\documentclass{INTERSPEECH2024}

\interspeechcameraready

\usepackage{multicol}
\usepackage{multirow}
\usepackage{color, colortbl}
\usepackage{xcolor}
\colorlet{eng}{blue!10}
\colorlet{cmn}{teal!10}
\colorlet{multi}{yellow!10}
\colorlet{euro}{orange!10}
\usepackage{pifont}
\newcommand{\cmark}{\ding{51}}%
\newcommand{\xmark}{\ding{55}}%
\usepackage{hyperref}
\usepackage{graphicx}
\usepackage{subcaption}
\usepackage{amsmath,amsfonts,bm}
\DeclareMathAlphabet{\mathsfit}{\encodingdefault}{\sfdefault}{m}{sl}
\SetMathAlphabet{\mathsfit}{bold}{\encodingdefault}{\sfdefault}{bx}{n}

\usepackage{tablefootnote}

\title{ML-SUPERB 2.0: Benchmarking Multilingual Speech Models \\ Across Modeling Constraints, Languages, and Datasets}
\name[affiliation={1}]{Jiatong}{Shi}
\name[affiliation={2,*}]{Shih-Heng}{Wang}
\name[affiliation={1,*}]{William}{Chen}
\name[affiliation={3,*}]{Martijn}{Bartelds}
\name[affiliation={1}]{Vanya Bannihatti}{Kumar}
\name[affiliation={1}]{Jinchuan}{Tian}
\name[affiliation={1}]{Xuankai}{Chang}
\name[affiliation={3}]{Dan}{Jurafsky}
\name[affiliation={3,4}]{Karen}{Livescu}
\name[affiliation={2}]{Hung-yi}{Lee}
\name[affiliation={1}]{Shinji}{Watanabe}
\address{ \thanks{* Equal contribution.}$^{1}$ Carnegie Mellon University, $^{2}$ National Taiwan University, $^{3}$ Stanford University, \\ $^{4}$ Toyota Technological Institute at Chicago}
\email{\{jiatongs, swatanab\}@cs.cmu.edu}

\usepackage[
backend=biber,
style=ieee,
citestyle=numeric-comp,
maxbibnames=3,
maxcitenames=3,
doi=false,isbn=false,url=false,eprint=false
]{biblatex}

\defbibheading{bibliography}[\refname]{}

\DeclareSourcemap{
	\maps[datatype=bibtex, overwrite=true]{
		\map{
			\step[fieldsource=booktitle,
			match=\regexp{.*Interspeech.*},
			replace={Proc. Interspeech}]
			\step[fieldsource=journal,
			match=\regexp{.*INTERSPEECH.*},
			replace={Proc. Interspeech}]
			\step[fieldsource=booktitle,
			match=\regexp{.*ICASSP.*},
			replace={Proc. ICASSP}]
			\step[fieldsource=booktitle,
			match=\regexp{.*icassp_inpress.*},
			replace={Proc. ICASSP (in press)}]
			\step[fieldsource=booktitle,
			match=\regexp{.*Acoustics,.*Speech.*and.*Signal.*Processing.*},
			replace={Proc. ICASSP}]
			\step[fieldsource=booktitle,
			match=\regexp{.*International.*Conference.*on.*Learning.*Representations.*},
			replace={Proc. ICLR}]
			\step[fieldsource=booktitle,
			match=\regexp{.*International.*Conference.*on.*Computational.*Linguistics.*},
			replace={Proc. COLING}]
			\step[fieldsource=booktitle,
			match=\regexp{.*SIGdial.*Meeting.*on.*Discourse.*and.*Dialogue.*},
			replace={Proc. SIGDIAL}]
			\step[fieldsource=booktitle,
			match=\regexp{.*International.*Conference.*on.*Machine.*Learning.*},
			replace={Proc. ICML}]
			\step[fieldsource=booktitle,
			match=\regexp{.*North.*American.*Chapter.*of.*the.*Association.*for.*Computational.*Linguistics:.*Human.*Language.*Technologies.*},
			replace={Proc. NAACL}]
			\step[fieldsource=booktitle,
			match=\regexp{.*Empirical.*Methods.*in.*Natural.*Language.*Processing.*},
			replace={Proc. EMNLP}]
			\step[fieldsource=booktitle,
			match=\regexp{.*Association.*for.*Computational.*Linguistics.*},
			replace={Proc. ACL}]
			\step[fieldsource=booktitle,
			match=\regexp{.*Automatic.*Speech.*Recognition.*and.*Understanding.*},
			replace={Proc. ASRU}]
			\step[fieldsource=booktitle,
			match=\regexp{.*Spoken.*Language.*Technology.*},
			replace={Proc. SLT}]
			\step[fieldsource=booktitle,
			match=\regexp{.*Speech.*Synthesis.*Workshop.*},
			replace={Proc. SSW}]
			\step[fieldsource=booktitle,
			match=\regexp{.*workshop.*on.*speech.*synthesis.*},
			replace={Proc. SSW}]
			\step[fieldsource=booktitle,
			match=\regexp{.*Advances.*in.*neural.*information.*processing.*},
			replace={Proc. NeurIPS}]
			\step[fieldsource=booktitle,
			match=\regexp{.*Advances.*in.*Neural.*Information.*Processing.*},
			replace={Proc. NeurIPS}]
			\step[fieldsource=booktitle,
			match=\regexp{.*Workshop.*on.* Applications.* of.* Signal.*Processing.*to.*Audio.*and.*Acoustics.*},
			replace={Proc. WASPAA}]
			\step[fieldsource=publisher,
			match=\regexp{.+},
			replace={{}}]
			\step[fieldsource=month,
			match=\regexp{.+},
			replace={{}}]
			\step[fieldsource=location,
			match=\regexp{.+},
			replace={{}}]
			\step[fieldsource=address,
			match=\regexp{.+},
			replace={{}}]
			\step[fieldsource=organization,
			match=\regexp{.+},
			replace={{}}]
			\step[fieldsource=pages,
			match=\regexp{.+},
			replace={{}}]
		}
	}
}


\begin{document}

\maketitle
 
\begin{abstract}
% 1000 characters. ASCII characters only. No citations.
% The Multilingual Speech Universal PERformance Benchmark (ML-SUPERB) extends the SUPERB benchmark into multilingual scenarios by including 143 languages. Following the SUPERB guidelines, 
ML-SUPERB evaluates self-supervised learning (SSL) models on the tasks of language identification and automatic speech recognition (ASR). This benchmark treats the models as feature extractors and uses a single shallow downstream model, which can be fine-tuned for a downstream task. However, real-world use cases may require different configurations. This paper presents ML-SUPERB~2.0, which is a new benchmark for evaluating pre-trained SSL and supervised speech models across downstream models, fine-tuning setups, and efficient model adaptation approaches. We find performance improvements over the setup of ML-SUPERB. However, performance depends on the downstream model design. Also, we find large performance differences between languages and datasets, suggesting the need for more targeted approaches to improve multilingual ASR performance.
% We envision the investigation as a pilot setup for a future multilingual benchmark.

\end{abstract}

\section{Introduction}
\label{sec: intro}

Modern multilingual speech models have the capacity to model hundreds or, in some cases, over a thousand languages~\cite{babu22_interspeech, pratap2023scaling, radford2023robust, peng2023reproducing, hou20_interspeech, pratap20c_interspeech, li22aa_interspeech, zhang2023google, chen2023improving, pratap2023scaling}, enabled by different training objectives, model architectures, and sources of training data.
%The capabilities of these models are enabled by various training objectives, model architectures, and sources of training data.
Importantly, the performance of these models is often evaluated using different experimental setups, which limits the extent to which their performance can be reliably compared.
%the performance of different models can be efficiently and robustly compared.
% This variation limits the extent to which the performance of different models can be efficiently and robustly compared.
Several 
%studies have tried to develop 
standardized evaluation setups and benchmarks have been proposed to evaluate the performance of pre-trained multilingual speech models~\cite{conneau22_interspeech, della2023cl, javed2023indicsuperb}.
%For example, XTREME-S~\cite{conneau22_interspeech} uses publicly available datasets to benchmark the performance of mSLAM~\cite{bapna2022mslam} and w2v-BERT~\cite{chung2021w2v} across several tasks, including automatic speech recognition (ASR), speech translation, and spoken language understanding.
% Another example is CL-MASR~\cite{della2023cl}, which evaluates multilingual speech models on the task of ASR.
% However, the benchmark specifically focuses on a continual learning setting.

The most comprehensive benchmark in terms of language coverage is the Multilingual Speech Universal PERformance Benchmark (ML-SUPERB)~\cite{mlsuperb}, which
%This benchmark 
covers 143 languages and includes multiple downstream tasks: monolingual ASR, multilingual ASR, and language identification (LID).
Like the original SUPERB~\cite{superb}, which only considers English speech, ML-SUPERB is set up to evaluate the performance of self-supervised learning (SSL) models.
This evaluation is performed by freezing their representations and treating the models as feature extractors.
These features are used as input to a lightweight downstream model, which can be fine-tuned for any of the downstream tasks.
To minimize the impact of the downstream model on the overall measured performance, a simple two-layer Transformer-based decoder is used.
% This decoder employs a connectionist temporal classification (CTC)~\cite{graves2006connectionist}.
ML-SUPERB was presented as a challenge at ASRU 2023, attracting 12 model submissions and 8 new language submissions~\cite{shi2023findings, yihuiasru, saktiasru, ogunremi2023r, chen2023evaluating, suwanbandit2023thai, sakti-2013, cahyawijaya-etal-2023-nusacrowd, effuse, chen2023joint, xue2023sshr}.
% This underscores its efficient and engaging setup to benchmark different SSL models.

Although the design of ML-SUPERB allows for efficient evaluation of multilingual SSL models across a large number of languages, it only considers one fixed downstream model design.
This is problematic, as past work has found that the choice of downstream model can affect the rankings of SSL models across downstream tasks~\cite{zaiem23b_interspeech,arora2024evaluation}. 
Also, the choice of downstream model designs can be affected by application requirements and users' budgets, which further motivates benchmarking with more flexible constraints.

In this paper, we present ML-SUPERB~2.0, which revisits ML-SUPERB's original design.
Specifically, ML-SUPERB~2.0 includes larger-scale downstream models, SSL model fine-tuning (including partial fine-tuning strategies), efficient pre-trained model adaptation techniques (adapters~\cite{houlsby2019parameter} and LoRA~\cite{hu2021lora}), and supervised pre-trained models (Whisper~\cite{radford2023robust} and OWSM~3.1~\cite{peng2024owsm}).
Also, we enrich ML-SUPERB's evaluation metrics to place greater focus on robustness across languages and describe variation across datasets.
%With ML-SUPERB~2.0, we aim for a more comprehensive and robust benchmark to promote multilingual speech processing research in the community.
All code and data used to develop ML-SUPERB~2.0 are publicly available.\footnote{\url{https://github.com/espnet/espnet/tree/master/egs2/ml_superb/asr1}}
% [ADDITIONAL RESULTS TO BE INCLUDED].

\section{Investigation Details}
\label{sec: methods}

% Building on the objectives outlined in Section~\ref{sec: intro}, this study aims to broaden the scope of ML-SUPERB by investigating four key areas: downstream architectures, model fine-tuning, efficient model adaptation, and the integration of supervised models. This section delves into the motivations and methodologies employed in each domain.

ML-SUPERB~2.0 considers a variety of architectural variations, pre-training and fine-tuning approaches, described in the next four sections.
We then discuss the changes in the evaluation metrics, which allow us to investigate performance differences across languages and datasets.

\subsection{Downstream Architectures}
\label{ssec: downstream}
% Prior research has delved into the impact of downstream architectures in various settings~\cite{chang2021exploration, zaiem23b_interspeech}. Chang et al.\cite{chang2021exploration} examined the interplay between CTC and hybrid CTC/attention architectures, employing Transformer or Conformer encoders\cite{watanabe2017hybrid, gulati20_interspeech, guo2021recent}. Zaiem et al.\cite{zaiem23b_interspeech} compared three architectures within the SUPERB framework, focusing on ASR tasks and evaluating performances across RNN-based CTC models, Conformer-based encoder-decoder structures, and ContextNet-based transducer frameworks~\cite{han20_interspeech}.

% These studies offer valuable insights into leveraging SSL representations in speech recognition. Nonetheless, the multilingual aspect has received limited attention. Our research extends these explorations within the ML-SUPERB framework, considering two architectures: a CTC-based encoder-only setup and a hybrid CTC/attention-based encoder-decoder configuration. We also evaluate three types of encoders: Transformer~\cite{vaswani2017attention}, Conformer~\cite{gulati20_interspeech}, and E-Branchformer~\cite{kim2023branchformer}.

Past work has found ASR performance differences between 
%frameworks and 
downstream architectures when comparing representations from pre-trained SSL models~\cite{chang2021exploration, zaiem23b_interspeech}.
% For example, performance differences have been found between the Transformer~\cite{vaswani2017attention} and Conformer~\cite{gulati20_interspeech} architectures.
% These differences occur within both CTC-based encoder-only (ENC) and hybrid CTC/attention-based encoder-decoder (ENC-DEC) frameworks~\cite{watanabe2017hybrid, chang2021exploration}.
% Also, performance differences have been observed when a bidirectional LSTM network~\cite{hochreiter1997}, Conformer, and a ContextNet transducer-based model~\cite{han20_interspeech} are compared.
These findings motivate a systematic comparison to better understand their impact on ASR performance. % when evaluating the ASR performance of pre-trained SSL models.
Therefore, ML-SUPERB~2.0 considers both CTC-based (CTC) and hybrid CTC/attention-based (CTC-ATT) frameworks as adopted in \cite{karita2019comparative, guo2021recent, peng2023comparative, zaiem23b_interspeech}, and within each framework, compares three architectures, namely the Transformer~\cite{vaswani2017attention}, Conformer~\cite{gulati20_interspeech}, and E-Branchformer~\cite{kim2023branchformer}.
In preliminary experiments, we compared these architectures to others (e.g., bi-LSTMs, transducers), and these three were chosen for their better performance or faster convergence.

\subsection{Model Fine-Tuning}
\label{ssec: finetuning}
% Fine-tuning pre-trained models is a prevalent practice in various domains, including multilingual speech processing. When the pre-training and fine-tuning tasks align, fine-tuning can be implemented directly without the need for additional adaptation networks~\cite{hou20_interspeech, radford2023robust, peng2023reproducing}. However, SSL or other unsupervised learning models often require supplementary downstream decoders, such as CTC-based or attention-based decoders for ASR tasks~\cite{baevski2020wav2vec, hsu2021hubert, shi2024multiresolution, babu22_interspeech, chang2021exploration}. Moreover, partial fine-tuning presents an efficient alternative, enabling a balance between training efficiency and effectiveness in target domains~\cite{wang2021fine, joshi2022simple, popuri22_interspeech, hou2021exploiting, chern2023audio}.

% In this context, our study introduces a fine-tuning dimension to ML-SUPERB, examining the combined impact of fine-tuning and downstream architectures. We also explore the efficacy of partial fine-tuning, concentrating on the top, middle, and bottom layers of the models to explore their performance impacts.

Fine-tuning is a common practice to adapt pre-trained SSL models to a downstream task.
% When the pre-training and fine-tuning tasks align, fine-tuning can be implemented directly without the need for training an additional downstream model~\cite{hou20_interspeech, radford2023robust, peng2023reproducing}.
% However, pre-trained SSL models, or other unsupervised learning models, often require downstream decoders to be fine-tuned on the task of ASR~\cite{baevski2020wav2vec, chang2021exploration, hsu2021hubert, babu22_interspeech, shi2024multiresolution}.
While fine-tuning is effective, it traditionally requires updating all model parameters, which is costly.
Partial fine-tuning is an alternative that strikes a balance between training efficiency and performance~\cite{kunze-etal-2017-transfer, hou2021exploiting}. % some original refs were not about ASR
ML-SUPERB~2.0 includes fine-tuning for the CTC/CTC-ATT frameworks, using either full fine-tuning or partial fine-tuning, which focuses on the bottom, middle, or top layers of the models, while keeping the other layers fixed.
% \footnote{For a 24-layer pre-trained model, the choices can be (1-6), (9-14), and (19-24) layers for partial fine-tuning.}

\subsection{Efficient Model Adaptation}
\label{ssec: adaptation}
% Efficient model adaptation offers a strategic approach to augmenting the performance of large-scale pre-trained models, providing a more resource-efficient alternative to full model fine-tuning~\cite{houlsby2019parameter, hu2021lora, lester2021power, li2021prefix}. These methods integrate lightweight modules into the pre-trained model, enabling targeted adjustments without the need to retrain the entire model. Such strategies have proven also beneficial in speech tasks, enhancing model performance, especially in low-resource contexts~\cite{hou2021exploiting, pham22_interspeech, thomas2022efficient, chen2023exploring, le2021lightweight}.

% In our study, we focus on two major efficient adaptation techniques: the Houlsby adapter~\cite{houlsby2019parameter} and low-rank adaptation~(LoRA)~\cite{hu2021lora}. The Houlsby adapter technique involves inserting two additional adapter layers into each Transformer block, with the rest of the blocks remaining unchanged. Conversely, LoRA incorporates learnable parameters into the matrix decomposition, specifically within the projection matrices of the multi-head attention mechanism. These adaptations are designed to refine the model's performance by introducing minimal yet impactful changes, facilitating a more efficient yet effective adaptation process for multilingual speech tasks.

Efficient model adaptation approaches offer a parameter-efficient alternative to full fine-tuning~\cite{houlsby2019parameter, hu2021lora, lester2021power, li2021prefix}.
In particular, the use of adapter models has been found to be competitive with, and sometimes improve upon, full fine-tuning, especially in low-resource settings~\cite{hou2021exploiting, pham22_interspeech, thomas2022efficient, chen2023exploring}.
These adapter models are small neural modules added between layers of a pre-trained model, which enable efficient fine-tuning by only learning the adapter module parameters.
ML-SUPERB~2.0 evaluates the performance of adapters using the CTC/CTC-ATT frameworks.
Specifically, we insert two adapter layers into each layer of the pre-trained SSL models, leaving the rest of the model unchanged (i.e.,~following the setup of~\cite{houlsby2019parameter}).
ML-SUPERB~2.0 also evaluates Low-Rank Adaptation~(LoRA).
LoRA freezes the pre-trained SSL models and injects low-dimensional layers to be added to the outputs of the projection matrices within the multi-head attention mechanism.

\subsection{Supervised Pre-Trained Models}
\label{ssec: supervised}
% Prior to the trend of using SSL models in speech processing, supervised models were the mainstay in the field. These models have also been effectively utilized as pre-trained models for various downstream tasks, demonstrating notable efficacy in multilingual speech recognition~\cite{hou20_interspeech, wu2021cross, cho2018multilingual}. Recent advancements in scaling up supervised models have further underscored their robustness and versatility, inspiring a range of studies to leverage pre-trained supervised models within specific domains~\cite{yeo2023visual, shao2023whisper, machavcek2023turning, zhuo2023lyricwhiz, barcovschi2023comparative, rouditchenko23_interspeech}.

% In our current research, we focus on evaluating two prominent supervised models: OpenAI's Whisper~\cite{radford2023robust} and OWSM 3.1~\cite{peng2023reproducing, peng2024owsm}. Whisper, developed by OpenAI, has been recognized for its robust performance in multilingual speech recognition, making it a compelling choice for our investigation. Similarly, OWSM 3.1 represents another significant advancement in supervised models, offering potential insights into the applicability and effectiveness of supervised pre-training. Through this exploration, we aim to assess how these supervised models can be integrated and utilized within the ML-SUPERB benchmark, contributing to a broader understanding of pre-training strategies for multilingual speech processing.

Scaling up supervised models has resulted in ASR performance that is competitive with SSL models on several evaluation datasets~\cite{radford2023robust, rouditchenko23_interspeech}.
ML-SUPERB~2.0 evaluates two recent supervised models, namely Whisper and OWSM~3.1, to relax the constraint of evaluating SSL models only.
We use the CTC framework to evaluate the encoder and the CTC-ATT framework to evaluate both the encoder and decoder of these models.
Also, we evaluate the partial fine-tuning setup described in Section~\ref{ssec: finetuning} within the CTC framework and use it exclusively within the CTC-ATT framework to limit the number of tunable parameters on the ML-SUPERB~2.0 dataset.

% While Whisper has been recognized for its robust performance in multilingual speech recognition, the model is not open-source.
% This could potentially limit a more detailed analysis on its performance beyond ML-SUPERB~2.0.
% Consequently, we also evaluate the performance of OWSM~3.1.
% This model improved over the original OWSM model~\cite{peng2023reproducing}, which aimed to reproduce Whisper-style training using publicly available data and an open-source toolkit.

% \subsection{Evaluation}
% \label{ssec: evaluation}
% For each configuration of the benchmark, ML-SUPERB~2.0 computes the mean performance and its standard deviation across languages.
% Inspired by past work on fairness in machine learning~\cite{pmlr-v80-hashimoto18a}, the benchmark also shows which languages are most negatively impacted by the frameworks and architectures evaluated.
% This is an attempt to refine the interpretation of the benchmark towards maximizing the performance of the worst-performing languages (see~\cite{rawls2001} for details).
% Lastly, we investigate performance differences between multiple evaluation sets in the same language, when available, in order to evaluate the impact of dataset characteristics and attempt to separate between the effects of language differences and domain/acoustic differences.

\section{Experimental Design}
\label{sec: exp}

% Building on the four configurations outlined in Section~\ref{sec: methods}, this section details the experimental setup of our study.

% Like ML-SUPERB, ML-SUPERB~2.0 evaluates a combined task that includes both multilingual ASR and language identification (LID).
% The objective is to concurrently predict the language ID token and transcribe the speech samples in our test dataset. 

ML-SUPERB~2.0 evaluates both multilingual ASR and LID.
The objective is to concurrently predict a language identifier token and transcribe the spoken content. ML-SUPERB~2.0 does not include ML-SUPERB's monolingual ASR track.

% \subsection{Data}
% \label{sec: data}

\subsection{General Setup}
\label{ssec: overall rule}

% Our investigation concentrates on a combined task that encompasses both multilingual ASR and language identification (LID). The objective is to concurrently predict the language ID and transcribe the spoken content. 

% We use the 1-hour benchmark dataset from ML-SUPERB, which allocates roughly one hour of speech data for each (language, corpus) pair.\footnote{The 1-hour ML-SUPERB dataset is a collection of 15 multilingual corpora, consisting of 143 languages. For each language in one of the source corpora, the benchmark select 1-hour training set, 10-minute development set, and 10-minute test set. } This dataset is in line with the ML-SUPERB framework, distinguishing between a set of languages for standard training and a few-shot language set.
% The first includes the full extent of available data, while the latter is limited to five utterances per language. Collectively, the dataset encompasses approximately 220 hours of training data, offering a broad spectrum of 143 languages and diverse domains for analysis. To facilitate a consistent and comparative evaluation, we adopt the same testing and development sets as those used in the ML-SUPERB benchmark, ensuring alignment with established benchmarks and enhancing the reliability of our findings.

ML-SUPERB~2.0 updates ML-SUPERB's dataset by correcting annotation mistakes,\footnote{We removed Highland Puebla Nahuatl from the Mexican endangered languages corpus and Norwegian from the NST corpus because of their mismatched annotations, and corrected the language label for VoxPopuli Italian.} resulting in $\sim$300 hours (85 hours for validation and test sets) drawn from 142 languages across 15 datasets.
% These datasets are similar to those used in 
% These corpora include Multilingual Librispeech \cite{pratap2020mls}, Commonvoice \cite{ardila2020common}, Voxforge \cite{maclean2018voxforge}, Voxpopuli \cite{wang2021voxpopuli}, Googlei18n open-source project \cite{kjartansson-etal-tts-sltu2018, kjartansson-etal-2020-open,he-etal-2020-open}, Nordic Language Technology ASR corpora \cite{rehm2012language}, Fleurs \cite{conneau2023fleurs}, NCHLT Speech \cite{barnard2014nchlt}, Spoken Wikipedia corpus \cite{baumann2019spoken}, Mexican endangered languages \cite{shi2021leveraging,shi2021highland,berrebbi22_interspeech}, M-AILab multilingual corpora \cite{mailab}, Living Audio dataset \cite{braude2019all}, and the ALFFA corpus \cite{de2014smartphone}.
%The data is from 143 languages across 15 datasets, and totals roughly 200 hours.
Some languages occur in more than one dataset. A 1-hour subset was drawn for each language-dataset pair, and the 1-hour subsets were combined to obtain the training dataset.
Similarly, 10-minute subsets were drawn for each language-dataset pair, and these serve as the development and test datasets.
% We ensure that there is no speaker overlap between the different datasets.
%From the train, development and test dataset, a 
A subset of 20 languages is reserved for few-shot (FS) learning experiments, whereas the normal experiments refer to other 122 languages.
%(i.e., the same 20 languages across the different data splits).
In the FS setting, five randomly selected utterances per language are used for training, while the 10-minute subsets for those languages are used for development and testing.

% In our study, we impose a limit of 100M tunable parameters during the training for our investigations, based on the following rationales:

% \begin{itemize}
%   \item Consistent with the objectives highlighted in Section~\ref{sec: intro}, this research aims to ease the restrictions initially set by ML-SUPERB. However, it's crucial to avoid potential biases that might arise from varying the number of tunable parameters across different experimental setups and the associated training efforts for each model.
%   \item We intend for our comparative analysis to contribute to the benchmark, offering insights that remain relevant with the progression towards larger pre-trained models. Setting a cap on tunable parameters facilitates a more equitable comparison with future models, aligning with the evolving trends towards larger model architectures.
%   \item Restricting the number of tunable parameters not only allow efficient training processes but also ensures that large-scale models can be evaluated across a diverse array of computing devices, enhancing the accessibility and practicality of the benchmarking process.
% \end{itemize}

%We benchmark both SSL and supervised models, and a
All experiments are performed using ESPnet~\cite{watanabe2018espnet} with SSL models support from S3PRL~\cite{superb}.
Among the SSL models available, we evaluate XLS-R~\cite{babu22_interspeech} and MMS~\cite{pratap2023scaling} 
%in ML-SUPERB~2.0 
due to their superior performance on ML-SUPERB.\footnote{We use model variants with 24 layers and $\sim$300 million parameters.} 
% We adopt the same ML-SUPERB strategies by using the layer-wise weighted summation of each layer from the SSL model, with a downstream module following.
As in ML-SUPERB, we compute a weighted sum of the layers of the SSL models and the encoder of the supervised models, and use it as input to the downstream models. This is applied to each of our experiments.
%Whisper and OWSM~3.1 are the evaluated supervised models.
% In the context, the MMS refers to the pre-trained wav2vec2 model.}
% For model decoding, we follow the experimental setup described in previous work \cite{karita2019comparative, guo2021recent, peng2023comparative}.

In line with the spirit of ML-SUPERB, ML-SUPERB~2.0 limits the number of tunable parameters to 100 million for each evaluated configuration.
This constraint ensures that large-scale models can be evaluated across a diverse range of computing environments, improving the accessibility and practicality of ML-SUPERB~2.0.

% In terms of decoding, the CTC framework uses a beam size of 10 across all configurations.
% For the CTC-ATT framework, the weight of CTC decoder is set to 0.3 and the weight of attention decoder is set to 0.7, respectively.
% This setup is in line with previous work~\cite{karita2019comparative, guo2021recent, peng2023comparative}.

\subsection{Downstream Architectures}
\label{ssec: exp downstream}
% As discussed in Sections \ref{ssec: downstream} and \ref{ssec: overall rule}, our analysis includes six downstream models, which encompass two model frameworks (ENC and ENC-DEC) and three network architectures. We choose hyperparameters informed by prior studies \cite{karita2019comparative, guo2021recent, peng2023comparative}, while maintaining fewer than 100M tunable parameters in each setting.

When evaluating the different architectures within the CTC and CTC-ATT frameworks, we base our hyperparameter selection on prior research~\cite{karita2019comparative, guo2021recent, peng2023comparative}. In particular, we keep the number of parameters of the downstream models below 100 million and tune only the learning rates.
For the CTC framework, the layer configurations are as follows: 24 layers for the Transformer-based model, 14 for the Conformer-based model, and 12 for the E-Branchformer-based model.
For the CTC-ATT models' encoders, we use 15 layers for the Transformer-based, 8 for the Conformer-based, and 7 for the E-Branchformer-based models.
The Conformer-based model has a kernel size of 15, whereas the E-Branchformer's multi-layer perceptron uses a kernel size of 31 and a dimension of 3072.
Common configurations across all models include an 8-head multi-head attention module with 512 hidden states and 2048 projection units, a batch size of 8 with gradient accumulation every four steps, a learning rate chosen from the range [$10^{-3}, 10^{-4}, 10^{-5}$] with 25,000 warm-up steps, and a dropout rate of 0.1.
For the decoders, a Transformer decoder with 8 layers is used for all models.
For hybrid training, the CTC and attention decoder weights are set to 0.3 and 0.7 respectively.
% This setup is in line with previous work~\cite{karita2019comparative, guo2021recent, peng2023comparative}.

\subsection{Model Fine-Tuning}
\label{ssec: exp fine-tune}
% As detailed in Section~\ref{ssec: finetuning}, our assessment of model fine-tuning consists of three partial fine-tuning approaches, specifically targeting the bottom (1-6), middle (9-14), and top layers (19-24) for 24-layer XLS-R and MMS. These approaches are designed to align with the threshold of 100M tunable parameters, as set forth in Section~\ref{ssec: overall rule}. Besides partial fine-tuning, we also examine the effects of full fine-tuning. %, which is provided solely for comparative analysis and does not adhere to the aforementioned guideline. 

ML-SUPERB~2.0 evaluates fine-tuning approaches using XLS-R and MMS, which both have 24 layers. The partial fine-tuning approach targets layers 1--6 (bottom), 9--14 (middle), or 19--24 (top).
This way, the number of updated parameters does not exceed 100 million.
Besides partial fine-tuning, we also examine full fine-tuning, which is provided only for comparison.
%solely as a comparative analysis.

To explore the impact of different downstream training objectives, we evaluate both the CTC and CTC-ATT frameworks. The CTC framework uses a 2-layer Transformer-encoder as in ML-SUPERB~\cite{mlsuperb}. For the CTC-ATT framework, we adopt a small-scale downstream model from the configuration in \cite{guo2021recent} to ensure that there are fewer than than 100 million tunable parameters. Specifically, the model consists of a 2-layer Transformer-based encoder and a 4-layer Transformer-based decoder. Each encoder block has an 8-head multi-head attention module with 256 hidden states and 1024 projection units, and each decoder block contains a 4-head multi-head attention module with 256 hidden states and 2048 linear projection units. The other hyperparameters are similar to those used for the experiments comparing downstream architectures.

\subsection{Efficient Model Adaptation}
% As outlined in Section~\ref{ssec: adaptation}, our experimentation includes the implementation of Housbly adapter and LoRA fine-tuning as methods for efficient data fine-tuning. We follow the detailed setting of two methods as \cite{chen2023exploring}: The Housbly adapter employs a bottleneck layer with a dimension of 64. In contrast, for LoRA fine-tuning, we assign a rank of 16 and set $\alpha$ to 16, integrating this module across all query and key vectors within the multi-head attention module of the SSL pre-trained model. The downstream analysis utilizes two Transformer-based frameworks: the ENC and ENC-DEC. The most of the hyperparameters are consistent with Section~\ref{ssec: exp downstream}. However, to offset the parameter increase from the adaptation layers, we decrease the encoder's layers by one.

We evaluate the use of adapters and LoRA within both frameworks and follow the setup described in Section~\ref{ssec: exp downstream}.
The configuration of the adapter models and LoRA follow previous work \cite{chen2023exploring}.
Specifically, the adapter layers have a dimension of 64, and we set the LoRA rank and its constant scaling factor $\alpha$ to 16.
The LoRA module is used across all query and key vectors within the multi-head attention module of the pre-trained SSL models.
To accommodate the additional parameters introduced by the adaptation layers, we reduce the number of layers in the encoder of the downstream models by one.

\subsection{Supervised Pre-Trained Models}
ML-SUPERB~2.0 evaluates the medium-sized variants of Whisper and OWSM 3.1, since these are 
%most closely aligned with the size of 
closest in size to the evaluated XLS-R and MMS models.\footnote{The Whisper and OWSM~3.1 model variants have 769 and 1017 million parameters, respectively.}
We include two experimental setups using these models, namely one using only their pre-trained encoder within the CTC framework, and another that evaluates both the pre-trained encoder and decoder within the CTC-ATT framework.
For the CTC framework, ML-SUPERB~2.0 investigates the performance of both the frozen pre-trained encoder using a Transformer-based downstream model and partial fine-tuning of the pre-trained encoder.
The experimental setup is similar to that for the CTC framework described in Sections~\ref{ssec: exp downstream} and \ref{ssec: exp fine-tune}, with the exception of fine-tuning only the top layers of the encoder (i.e., layers 19-24) to limit the number of updated parameters to 100 million.
In the CTC-ATT framework, we do not add additional downstream models. The encoder remains frozen and we also use the same settings (i.e., medium-sized model variant) as in the CTC framework.
Moreover, fine-tuning only targets the top layers of the decoder (i.e., layers 19-24).

% We use Whisper and OWSM 3.1 for the investigation of using supervised pre-trained model. We select the medium-size for both models. For each model, we conduct two sets of experiments, including one using only the encoder with a downstream model and the other using both its own encoder and decoder. For experiments in the first set, we test two cases including the use of frozen pre-trained encoder and a large downstream CTC model and pre-trained encoder fine-tuning wiht a light downstream CTC model. For experiments with the frozen encoder, we use the same architecture as the Transformer-based CTC model in downstream experiments. For fine-tuning experiments, we only conduct partial fine-tuning on the top layers. For encoder-only model, we use the top layers in the encoder. For encoder-decoder model, we use the top layers in the decoder.

\subsection{Evaluation}
% To compute an aggregated CER for each evaluated configuration, ML-SUPERB~2.0 first calculates the CER for each individual language-dataset pair.
% Subsequently, the macro-average of the CERs is computed across all datasets of each language.
% This procedure results in language-specific CERs, from which the final macro-averaged CER and associated standard deviation are computed.
% Moreover, the language-specific CERs are used to report the worst-performing language for each evaluated configuration.

For each configuration of the benchmark, ML-SUPERB~2.0 computes the LID accuracy and character error rates (CER) on the test dataset.
Specifically, we first compute a per-language CER as the macro-average of CERs across all of the (one or more) datasets per language.  We then compute
the macro-average of the per-language CERs
%(of the macro-average across datasets) 
and the standard deviation (SD) of the language-specific CERs.  We report these
%are computed based on these language-specific CERs, which we report 
for both the normal and few-shot~(FS) settings. The LID accuracy scores are only reported for the normal setting.
Inspired by past work on fairness in machine learning~\cite{pmlr-v80-hashimoto18a}, we also report the worst-performing language (WL), i.e.~the one with the highest CER in the normal setting, for each configuration, in an attempt to encourage research on methods that leave no language "behind".
%ensure that no language is ``left behind".
%This is an attempt to understand how the different frameworks and architectures impact the performance of the worst-performing languages.
% Lastly, we report the CER range between multiple datasets in the same language, when available, to separate between the effects of domain or acoustic differences. We perform this analysis using the best-performing model for each experiment mentioned in the previous subsections given the CER in the normal setting.
% We report the language that shows the highest range in CER among its datasets.
Lastly, we investigate the CER range between multiple datasets in the same language, when available, to separate the effects of domain or acoustic differences. We perform this analysis using the best-performing model and configuration of the benchmark given the CER in the normal setting.
We describe the language that shows the highest range in CER among its datasets.

\begin{table}[!t]
    \centering
    % \caption{Investigation on downstream architectures. The number of tunable parameters are in the parentheses of the ``Param" column. + indicate the use of encoder-decoder architecture. The few-shot results are in ``CER (FS)" column. }
    \caption{Results of the downstream architecture experiments, showing the downstream model, number of model parameters (tunable parameters in parentheses), LID accuracy (ACC), aggregated CERs and few-shot CERs (FS) with standard deviations, and CERs for the worst-performing language (WL). T., C., E-B. are abbreviations for Transformer, Conformer, and E-Branchformer. +~indicates the use of the CTC-ATT framework. $\dagger$ refers to the original ML-SUPERB setting~\cite{mlsuperb}.}
    \vspace{-10pt}
    \resizebox{\linewidth}{!}{
    \begin{tabular}{l|l|c|c|ccc} 
        \toprule
        \multirow{2}{*}{Models} & \multirow{2}{*}{Method} & \multirow{2}{*}{Param. (M)} & \multirow{2}{*}{ACC} & \multicolumn{3}{c}{CER}  \\
        \cmidrule{5-7}
         & & & & Normal & FS & WL \\
        \midrule
        XLS-R  & T.\textsuperscript{$\dagger$}        & 323.7 (\phantom{0}6.3) & 90.9 & 24.8 $\pm$ 12.1 & 34.4 $\pm$ 21.1 & \phantom{0}75.1 \\
        MMS    & T.\textsuperscript{$\dagger$}        & 321.8 (\phantom{0}6.3) & 90.3 & 24.7 $\pm$ 12.3 & \textbf{31.0} $\pm$ 18.6 & \phantom{0}67.6  \\
        \midrule
        \multirow{6}{*}{XLS-R } & T.        & 408.5 (91.1) & 93.7 & 20.7 $\pm$ 10.8 & 33.3 $\pm$ 20.8 & \phantom{0}68.0 \\
                     &  C.                   & 408.9 (91.5) & 82.3 & 22.9 $\pm$ 12.8 & 33.4 $\pm$ 20.5 & \phantom{0}86.9 \\
                     &  E-B.                     & 409.6 (92.2) & 94.1 & 18.2 $\pm$ 10.6 & 32.3 $\pm$ 20.9 & \phantom{0}69.5 \\
                     \cmidrule{2-7}
                     & T.\textsuperscript{+}    & 416.0 (98.6) & 93.6  & 19.2 $\pm$ 11.9 & 33.6 $\pm$ 21.0 & \phantom{0}76.2 \\
                     & C.\textsuperscript{+}   & 416.3 (98.9) & 83.7 & 23.9 $\pm$ 19.1 & 34.8 $\pm$ 22.6 & 102.9  \\
                     & E-B.\textsuperscript{+}   & 417.1 (99.7)  &  94.7 & 16.9 $\pm$ 10.7 & 32.3 $\pm$ 21.1 & \phantom{0}63.8 \\
        \midrule
        \multirow{6}{*}{MMS}  & T.        & 406.6 (91.1) & 93.6 & 21.0 $\pm$ 11.2 & 31.7 $\pm$ 19.3 & \phantom{0}67.4 \\
                     &  C.                  &  407.0 (91.5) & 85.3 & 22.7 $\pm$ 14.2 & 31.7 $\pm$ 17.7 & \phantom{0}94.6 \\
                     &  E-B.             & 407.7 (92.2) & 93.0 & 20.4 $\pm$ 10.6 & \textbf{31.0} $\pm$ 19.1 & \textbf{\phantom{0}61.5} \\
                     \cmidrule{2-7}
                     & T.\textsuperscript{+}    & 414.1 (98.6) & 94.3 & 18.8 $\pm$ 11.8 & 31.9 $\pm$ 19.0 & \phantom{0}73.1 \\
                     & C.\textsuperscript{+}   & 414.4 (98.9) & 84.0 & 23.8 $\pm$ 16.7 & 33.6 $\pm$ 18.5 & 106.1 \\
                     & E-B.\textsuperscript{+}   & 415.2 (99.7)  & \textbf{95.2} & \textbf{16.6} $\pm$ 11.8 & 32.6 $\pm$ 20.4 & \phantom{0}69.8 \\
        \bottomrule
    \end{tabular}
    }
    \vspace{-10pt}
    \label{tab:downstream}
\end{table}
\begin{table}[!t]
    \centering
    % \caption{Investigation on pre-trained model fine-tuning. The number of tunable parameters are in the parentheses of the ``Param" column. + indicate the use of encoder-decoder architecture. The few-shot results are in ``CER (FS)" column.}
    \caption{Results of the fine-tuning experiments, showing the method, number of model parameters (tunable parameters in parentheses), LID accuracy (ACC), aggregated CERs and few-shot CERs (FS) with standard deviations, and CERs for the worst-performing language (WL). +~indicate the use of the CTC-ATT framework. $\dagger$ refers to the original ML-SUPERB setting~\cite{mlsuperb}.}
    \vspace{-10pt}
    \resizebox{\linewidth}{!}{
    \begin{tabular}{l|l|c|c|ccc} 
        \toprule
        \multirow{2}{*}{Models} & \multirow{2}{*}{Method} & \multirow{2}{*}{Param. (M)} & \multirow{2}{*}{ACC} & \multicolumn{3}{c}{CER}  \\
        \cmidrule{5-7}
         & & & & Normal & FS & WL \\
        \midrule
        XLS-R  & -\textsuperscript{$\dagger$}        & 323.7 (\phantom{0}\phantom{0}6.3) & 90.9 & 24.8 $\pm$ 12.1 & 34.4 $\pm$ 21.1 & \phantom{0}75.1  \\
        MMS    & -\textsuperscript{$\dagger$}        & 321.8 (\phantom{0}\phantom{0}6.3) & 90.3 &  24.7 $\pm$ 12.3 & 31.0 $\pm$ 18.6 & \phantom{0}67.6 \\
        \midrule
        \multirow{8}{*}{XLS-R } &  1-6      & 323.7 (\phantom{0}90.3) & 91.7 & 20.5 $\pm$ 12.8 & 29.4 $\pm$ 17.8 & \phantom{0}74.0 \\
                     &  9-14     & 323.7 (\phantom{0}90.3) & 93.0 & 18.5 $\pm$ 12.8 & 31.3 $\pm$ 21.3 & \phantom{0}73.2 \\
                     &  19-24    & 323.7 (\phantom{0}90.3) & 91.4 & 22.0 $\pm$ 13.2 & 31.8 $\pm$ 20.8 & \phantom{0}74.8 \\
                     & 1-24 & 323.7 (323.7) & 94.3 & 15.8 $\pm$ 12.4 & 28.6 $\pm$ 20.2 & \phantom{0}70.2 \\
                     \cmidrule{2-7}
                     &  1-6\textsuperscript{+}      & 333.4 (\phantom{0}99.9) & 84.0 & 30.5 $\pm$ 22.8 & 35.4 $\pm$ 18.1 & 119.1 \\
                     &  9-14\textsuperscript{+}      & 333.4 (\phantom{0}99.9) & 93.2 & 22.7 $\pm$ 18.3 & 32.2 $\pm$ 18.7 & \phantom{0}96.0 \\
                     &  19-24\textsuperscript{+}     & 333.4 (\phantom{0}99.9) & 89.8 & 25.6 $\pm$ 19.5 & 32.2 $\pm$ 18.4 & 101.5 \\
                     & 1-24\textsuperscript{+}   & 333.4 (333.4) & 94.1 &  16.8 $\pm$ 14.3 & 29.5 $\pm$ 17.2 & \phantom{0}79.0  \\
        \midrule
        \multirow{8}{*}{MMS}  &  1-6      & 321.8 (\phantom{0}90.8) & 93.8 &  18.8 $\pm$ 12.0 & 31.0 $\pm$ 20.8 & \phantom{0}75.6 \\
                     &  9-14                   & 321.8 (\phantom{0}90.8) &  95.6 & \textbf{15.5} $\pm$ 10.3 & \textbf{27.7} $\pm$ 16.7 & \textbf{\phantom{0}62.7} \\
                     &  19-24                   & 321.8 (\phantom{0}90.8) & 93.4 & 19.4 $\pm$ 14.6 & 28.5 $\pm$ 17.8 & \phantom{0}96.2 \\
                     & 1-24 & 321.8 (321.8)& 87.7 & 27.4 $\pm$ 13.6 & 31.7 $\pm$ 18.8 & \phantom{0}80.5 \\
                     \cmidrule{2-7}
                     &  1-6\textsuperscript{+}       & 331.4 (100.5) & 93.6 & 25.4 $\pm$ 16.4 & 35.9 $\pm$ 19.6 & \phantom{0}91.2 \\
                     &  9-14\textsuperscript{+}     & 331.4 (100.5) & \textbf{95.7} & 17.6 $\pm$ 14.6 & 28.9 $\pm$ 16.8 & \phantom{0}89.5  \\
                     &  19-24\textsuperscript{+}    & 331.4 (100.5) & 92.1 & 23.2 $\pm$ 21.6 & 28.5 $\pm$ 17.5 & 119.7 \\
                     & 1-24\textsuperscript{+}  & 331.4 (331.4) & 95.5 & 15.9 $\pm$ 15.0 & 30.2 $\pm$ 20.7 & \phantom{0}81.6 \\
        \bottomrule
    \end{tabular}
    }
    \vspace{-15pt}
    \label{tab:fine-tuning}
\end{table}

\section{Results and Discussion}

\subsection{Comparisons Between Models and Settings}

\noindent \textbf{Downstream Architectures}: The results for different downstream architectures are presented in Table~\ref{tab:downstream}.
The table shows that there is no superior model across all evaluated configurations.
However, the E-Branchformer-based models outperform their Transformer-based and Conformer-based counterparts in almost all cases.
This result aligns with trends noted in previous work~\cite{peng2023comparative}, confirming the strong performance of the E-Branchformer model for LID and multilingual ASR.

When comparing the CTC and CTC-ATT frameworks, we find that CTC performs slightly better in the few-shot setting, while CTC-ATT (i.e, rows with a plus) is stronger in the normal setting. The findings suggest that the CTC framework might have better generalization capabilities when limited amounts of data are available. Comparing these results to the shallow-downstream baseline from ML-SUPERB (i.e., first two rows), we find an improvement in LID and ASR performance in the normal setting. However, the shallow-downstream baseline, based on MMS, still performs competitively in the few-shot setting.
With roughly 6 million tunable parameters, the baseline's performance echos the insight from the 2023 ML-SUPERB challenge \cite{shi2023findings}: scaling up models does not necessarily translate to improved performance on multilingual speech tasks.

In sum, our results reinforce findings in past work \cite{zaiem23b_interspeech} that pre-trained SSL model rankings for ASR vary with the choice of downstream architecture.

\begin{table}[!t]
    \centering
    % \caption{Investigation on efficient model adaptation. The number of tunable parameters are in the parentheses of the ``Param" column. + indicate the use of encoder-decoder architecture. The few-shot results are in ``CER (FS)" column.}
    \caption{Results of the efficient model adaptation experiments, showing the method, number of model parameters (tunable parameters in parentheses), LID accuracy (ACC), aggregated CERs and few-shot CERs (FS) with standard deviations, and CERs for the worst-performing language (WL). +~indicate the use of the CTC-ATT framework. $\dagger$ refers to the original ML-SUPERB setting~\cite{mlsuperb}.}
    \vspace{-10pt}
    \resizebox{\linewidth}{!}{
    \begin{tabular}{l|l|c|c|ccc} 
        \toprule
        \multirow{2}{*}{Models} & \multirow{2}{*}{Method} & \multirow{2}{*}{Param. (M)} & \multirow{2}{*}{ACC} & \multicolumn{3}{c}{CER}  \\
        \cmidrule{5-7}
         & & & & Normal & FS & WL \\
        \midrule
        XLS-R  & -\textsuperscript{$\dagger$}        & 323.7 (\phantom{0}\phantom{0}6.3) & 90.9 & 24.8 $\pm$ 12.1 & 34.4 $\pm$ 21.1 & 75.1  \\
        MMS    & -\textsuperscript{$\dagger$}        & 321.8 (\phantom{0}\phantom{0}6.3) & 90.3 & 24.7 $\pm$ 12.3 & \textbf{31.0} $\pm$ 18.6 & 67.6  \\
        \midrule
        \multirow{4}{*}{XLS-R } &  LoRA      & 410.1 (\phantom{0}92.7) & \textbf{94.4} & 20.3 $\pm$ 10.7 & 33.2 $\pm$ 21.2 & \textbf{63.0} \\
                     &  Adapter                    & 411.7 (\phantom{0}94.3) & 94.2 & 20.6 $\pm$ 10.8 & 33.7 $\pm$ 20.8 & 67.3 \\
                     \cmidrule{2-7}
                     &  LoRA\textsuperscript{+}  &   415.8 (\phantom{0}98.4)   &  93.8 &  19.1 $\pm$ 11.9 & 33.7 $\pm$ 20.7 & 69.7 \\
                     &  Adapter\textsuperscript{+}  &  417.4 (100.0) & 93.4 & 19.5 $\pm$ 11.7 & 33.3 $\pm$ 20.8 & 72.3 \\
        \midrule
        \multirow{4}{*}{MMS}  &   LoRA             & 408.2 (\phantom{0}92.7) & 93.5 & 21.3 $\pm$ 10.9 & 31.5 $\pm$ 18.3 & 65.8 \\
                     &   Adapter          & 409.8 (\phantom{0}94.3) & 91.7 & 24.5 $\pm$ 11.0 & 35.5 $\pm$ 18.9 & 70.6 \\
                     \cmidrule{2-7}
                     &   LoRA\textsuperscript{+}  &   413.7 (\phantom{0}98.4)   & 94.2 & \textbf{18.7} $\pm$ 11.5 & 32.6 $\pm$ 20.0 & 68.0 \\
                     &   Adapter\textsuperscript{+}  &  415.5 (100.0) & 92.3 & 21.9 $\pm$ 12.2 & 35.7 $\pm$ 19.5 & 77.2 \\
        \bottomrule
    \end{tabular}
    }
    \vspace{-10pt}
    \label{tab:adaptation}
\end{table}
\begin{table}[!t]
    \centering
    % \caption{Investigation on using supervised pre-trained model. The number of tunable parameters are in the parentheses of the ``Param" column. * indicate that only encoder of the supervised model is used, while the other cases use both encoder and decoder. The few-shot results are in ``CER (FS)" column.}
    \caption{Results of the supervised model experiments, showing whether fine-tuning (FT) is performed, number of model parameters (tunable parameters in parentheses), LID accuracy (ACC), aggregated CERs and few-shot CERs (FS) with standard deviations, and CERs for the worst-performing language (WL). Asterisks indicate that only the encoder is used. $\dagger$ refers to the original ML-SUPERB setting~\cite{mlsuperb}.}
    \vspace{-10pt}
    \resizebox{\linewidth}{!}{
    \begin{tabular}{l|l|c|c|ccc} 
        \toprule
        \multirow{2}{*}{Models} & \multirow{2}{*}{FT} & \multirow{2}{*}{Param. (M)} & \multirow{2}{*}{ACC} & \multicolumn{3}{c}{CER}  \\
        \cmidrule{5-7}
         & & & & Normal & FS & WL \\
        \midrule
        XLS-R  & \xmark\textsuperscript{$\dagger$}        & \phantom{0}323.7 (\phantom{0}\phantom{0}6.3) & 90.9 & 24.8 $\pm$ 12.1 & 34.4 $\pm$ 21.1 & \phantom{0}75.1 \\
        MMS    & \xmark\textsuperscript{$\dagger$}        & \phantom{0}321.8 (\phantom{0}\phantom{0}6.3) & 90.3 & 24.7 $\pm$ 12.3 & 31.0 $\pm$ 18.6 & \textbf{\phantom{0}67.6}  \\
        \midrule
        \multirow{3}{*}{Whisper } &  \xmark*  & \phantom{0}515.8 (\phantom{0}91.1) & \textbf{91.7} & \textbf{21.0} $\pm$ 12.5 & \textbf{27.4} $\pm$ 13.3 & \phantom{0}82.9 \\
                     &  \cmark*      & \phantom{0}431.0 (\phantom{0}90.7) & 83.9 & 26.8 $\pm$ 15.0 & 29.6 $\pm$ 13.5 & \phantom{0}93.5 \\
                     % &  Lora*      &  & \\
                     % &  Housbly*                    &  &  & \\
                     \cmidrule{2-7}
                     &  \cmark  & \phantom{0}762.3 (\phantom{0}84.4) & 85.5 & 25.6 $\pm$ 19.4 & 35.0 $\pm$ 17.5 & 107.2 \\
                     % &  Lora &      &  & \\
                     % &  Housbly &   &  & \\
        \midrule
        \multirow{3}{*}{OWSM}  &  \xmark*  & \phantom{0}671.2 (\phantom{0}88.4) & 77.8 & 27.8 $\pm$ 22.6 & 31.7 $\pm$ 17.3 & \phantom{0}99.9 \\
                     &  \cmark*      & \phantom{0}612.1 (\phantom{0}88.4) & 71.0 & 24.9 $\pm$ 14.9 & 31.5 $\pm$ 16.9 & \phantom{0}99.7 \\
                     % &  Lora*      &  & \\
                     % &  Housbly*                    &  &  & \\
                     \cmidrule{2-7}
                     &  \cmark  & 1016.9 (100.8) & 80.5 & 40.0 $\pm$ 41.8 & 40.0 $\pm$ 24.9 & 337.6 \\
                     % &  Lora &      &  & \\
                     % &  Housbly &   &  & \\
        \bottomrule
    \end{tabular}
    }
    \vspace{-10pt}
    \label{tab:supervised}
\end{table}

\noindent \textbf{Model Fine-tuning}: 
The model fine-tuning results are presented in Table~\ref{tab:fine-tuning}.
These results suggest that fine-tuning of the middle layers (i.e.,~layers 9–14) is more effective across the evaluated SSL models and training frameworks than %the other partial fine-tuning strategies (i.e.,~ 
fine-tuning the bottom or top layers.
% Particularly within the MMS-based CTC framework, fine-tuning the middle layers significantly outperforms fine-tuning the entire SSL model.
While full fine-tuning mostly outperforms partial fine-tuning in the normal setting (it also has the lowest mean CER on the worst-performing language in most cases), this is not the case in the FS setting.
For instance, full fine-tuning of MMS leads to a higher mean CER compared to fine-tuning the middle layers in the FS setting.
This suggests that the choice of fine-tuning strategy is crucial and warrants further exploration within the context of the benchmark.

\noindent \textbf{Efficient Model Adaptation}: The efficient model adaptation results, detailed in Table~\ref{tab:adaptation}, also do not reveal a single best model across the evaluated configurations.
However, LoRA outperforms adapters across SSL models in the normal setting, indicating it is the preferred option within the setup of the benchmark.
When comparing frameworks, the results generally align with those from the downstream analysis~(Table~\ref{tab:downstream}). 
We find a difference when looking at the LID task, where XLS-R with LoRA adaptation outperforms MMS within the CTC framework, while MMS achieves better performance within the CTC-ATT framework. This suggests that the choice of framework and adaptation method can impact the performance, depending on the task and the SSL model used.

\noindent \textbf{Supervised Pre-Trained Models}: The experiments with supervised pre-trained models are shown in Table~\ref{tab:supervised}. The results indicate that using only the pre-trained encoder from supervised models leads to better ASR performance than using models with the original decoder. The performance differences might stem from challenges in partial fine-tuning of the decoder, or from the potential biases from large-scale supervised training in major languages.
Also, we find that supervised pre-trained models do not consistently outperform the SSL-based models across the evaluated configurations, which aligns with results reported in previous work~\cite{rouditchenko23_interspeech}.
While this work does not conduct a deeper analysis into the optimal utilization of supervised pre-trained models, it highlights this area as a promising direction for future research within the ML-SUPERB 2.0 benchmark.
% The inclusion of supervised models in ML-SUPERB~2.0 aims to spur further exploration into how supervised pre-trained models can be effectively applied across a broader spectrum of languages.

\subsection{Variation Across Languages and Datasets}
% Performance differences between different languages and their datasets (by Martijn)

To investigate the impact of different languages on the benchmark performance, we report a standard deviation for each reported CER.
We find large standard deviations in both the normal and few-shot settings, indicating that there is substantial variation among the language-specific CERs.
The CER of the worst-performing language, which we found to be Lao or Min Nan Chinese in most cases, also highlights the large impact of language differences, since it is substantially higher than the mean CER in the normal and few-shot settings.
% , as indicated by the large standard deviations in the normal and few-shot settings.
% This observation applies to each of the evaluated downstream architectures, fine-tuning methods, efficient model adaptation approaches, and is also observed for the supervised models.

When investigating performance differences between datasets within a single language, we find large differences as well. 
% Particularly, Urdu and Irish, which have data from Fleurs~\cite{conneau2023fleurs}, Common Voice~\cite{ardila2020common}, and the Living Audio Dataset~\cite{braude2019all}, show the largest range in CER across their datasets.
% Specifically, the CER ranged between 29.9\% and 76.7\% for Urdu given the downstream architecture experiments, between 9.4\% and 35.1\% for Irish given the fine-tuning experiments, between 31.2\% and 84.3\% for Urdu given the efficient model adaptation approaches, and between 46.3\% and 79.5\% for Urdu when using the best-performing supervised model setup.
% Particularly, we find large differences between the datasets of Urdu across a large number of evaluated configurations.
% For example, when we investigate the results of the best-performing model and configuration of ML-SUPERB~2.0 (i.e.~fine-tuning the middle layers of MMS within the CTC framework), we find that the CER of Urdu on Common Voice~\cite{ardila2020common} is 21.8\%, while the CER on Fleurs~\cite{conneau2023fleurs} is 56.9\%.
For the best-performing model and configuration of ML-SUPERB~2.0, which involves fine-tuning the middle layers of MMS within the CTC framework, the largest differences in CER are among the datasets of Urdu.
Specifically, we find that the CER of Urdu from Common Voice~\cite{ardila2020common} is 21.8\%, whereas it is 56.9\% on data from Fleurs~\cite{conneau2023fleurs}.
Note also that Urdu has the largest performance difference between its datasets in many of the other evaluated configurations.

% Urdu and Irish have data from Fleurs~\cite{conneau2023fleurs}, Common Voice~\cite{ardila2020common}, and the Living Audio Dataset~\cite{braude2019all}.
% In particular, we find some of the largest standard deviations when either the bottom or top layers of XLS-R are fine-tuned within the CTC-ATT framework.

These results motivate future work on creating truly multilingual model representations, which can transfer to a broad range of languages and domains.

% Ranges:
% Downstream:
%     Language Dataset_Max  Max Sum/Avg Dataset_Min  Min Sum/Avg  Range
% 115      urd      fleurs         76.7          cv         29.9   46.8
% Fine-tuning:
%     Language Dataset_Max  Max Sum/Avg Dataset_Min  Min Sum/Avg  Range
%    Language Dataset_Max  Max Sum/Avg Dataset_Min  Min Sum/Avg  Range
% 31      gle      fleurs         35.1         lad          9.4   25.7
% Eff mod adapt:
%     Language Dataset_Max  Max Sum/Avg Dataset_Min  Min Sum/Avg  Range
% 115      urd      fleurs         84.3          cv         31.2   53.1
% Sup mods:
%     Language Dataset_Max  Max Sum/Avg Dataset_Min  Min Sum/Avg  Range
% 115      urd      fleurs         79.5          cv         46.3   33.2

% \begin{enumerate}
%     \item Compute dataset range for simplest model (original ML-SUPERB; XLS-R first row in table 1)
%     \begin{enumerate}
%         \item XLSR: urd. min: (cv, 27.5) max (fleurs, 68.9) range 41.4.
%         \item MMS: urd. min: (cv, 30.2) max (fleurs, 76.9) range 46.7.
%     \end{enumerate}
%     \item Compute dataset range for best-performing downstream model approach (i.e., from table 1; XLSR + E-branchformer)
%     \item Compute dataset range for best-performing model given fine-tuning setting (i.e., from table 2 or 3; full fine-tuning of XLS-R).
%     \item efficient model adapt.
%     \item Compute dataset range for best-performing model given supervised models (i.e., table 4; TBD).
% \end{enumerate}

% \subsection{Further Discussion}

\section{Conclusion}

% Perhaps something should be written about the large standard deviations, and that language-specific strategies might be needed to minimize the variability in performance.

We introduced ML-SUPERB~2.0, an updated benchmark for multilingual speech pre-trained models, which builds upon and extends ML-SUPERB. By relaxing many of ML-SUPERB's constraints, ML-SUPERB 2.0 opens up new avenues for research, offering a broader scope for exploration within the benchmark's setup. We investigated four primary extensions to ML-SUPERB, namely the use of larger-scale downstream models, model fine-tuning, efficient model adaptation, and the incorporation of supervised pre-trained models. Furthermore, we enhanced the evaluation metrics of ML-SUPERB to better track robustness across languages, and described dataset variation using the benchmark's best-performing model and configuration.

% Through comprehensive experimentation, we provide an initial assessment of these directions.
While each of the four extensions has shown improvements over the models in the original ML-SUPERB, model fine-tuning achieves the best performance on both LID and multilingual ASR.
% However, the models and setup evaluated in ML-SUPERB still shows the lowest CER for the worst-performing language.
However, the large deviations across languages and the substantially higher CER for the worst-performing languages suggest that tailored or language-specific approaches might be essential to reduce performance variability and improve model efficacy in multilingual speech processing.

\section{Acknowledgements}
This work used the Bridges2 system at PSC and Delta system at NCSA through allocations CIS210014 and IRI120008P from the ACCESS program, supported by NSF grants \#2138259, \#2138286, \#2138307, \#2137603, and \#2138296.
% \ifinterspeechfinal
%      The INTERSPEECH 2023 organisers
% \else
%      The authors
% \fi
% would like to thank ISCA and the organising committees of past INTERSPEECH conferences for their help and for kindly providing the previous version of this template.

% \bibliographystyle{IEEEtran}
% \bibliography{mybib}

\section{References}
{
\printbibliography
}

\end{document}